\documentstyle[10pt,epsf,dp_delphititle]{dp_delphi}
\makeindex
\def\DpPaperGroup{EP}
\def\DpPaperRef{2000-089}
\def\DpDate{Revised 30 October 2000}
\def\DpAuthors{DELPHI Collaboration}
\def\DpSubmit{(Submitted to Phys.Lett.B)}
\def\DpTitle{{Measurement of the $ZZ$ cross-section in $e^+ e^-$ interactions
at 183-189 GeV}}
\def\DpComment{ }
\def\DpEMail{ }

\newcommand{\z}{$Z$~}
\newcommand{\zz}{$ZZ$~}
\newcommand{\zg}{$Z\gamma^*$~}
\newcommand{\ww}{$WW$~}
\newcommand{\qcd}{q$\rm{\overline q}(\gamma)$~}
\newcommand{\qqqq}{$\rm{ q \overline q q \overline q}$~}
\newcommand{\ba}{\begin{array}}
\newcommand{\ea}{\end{array}}
\newcommand{\bc}{\begin{center}}
\newcommand{\ec}{\end{center}}
\newcommand{\bt}{\begin{tabular}}
\newcommand{\et}{\end{tabular}}
\newcommand{\beq}{\begin{eqnarray}}
\newcommand{\eeq}{\end{eqnarray}}
\newcommand{\bes}{\begin{eqnarray*}}
\newcommand{\ees}{\end{eqnarray*}}
\include{qqnn_dn_definitions}
\begin{document}
\makeatletter
\newcount\@tempcntc
\def\@citex[#1]#2{\if@filesw\immediate\write\@auxout{\string\citation{#2}}\fi
  \@tempcnta\z@\@tempcntb\m@ne\def\@citea{}\@cite{\@for\@citeb:=#2\do
    {\@ifundefined
       {b@\@citeb}{\@citeo\@tempcntb\m@ne\@citea\def\@citea{,}{\bf ?}\@warning
       {Citation `\@citeb' on page \thepage \space undefined}}%
    {\setbox\z@\hbox{\global\@tempcntc0\csname b@\@citeb\endcsname\relax}%
     \ifnum\@tempcntc=\z@ \@citeo\@tempcntb\m@ne
       \@citea\def\@citea{,}\hbox{\csname b@\@citeb\endcsname}%
     \else
      \advance\@tempcntb\@ne
      \ifnum\@tempcntb=\@tempcntc
      \else\advance\@tempcntb\m@ne\@citeo
      \@tempcnta\@tempcntc\@tempcntb\@tempcntc\fi\fi}}\@citeo}{#1}}
\def\@citeo{\ifnum\@tempcnta>\@tempcntb\else\@citea\def\@citea{,}%
  \ifnum\@tempcnta=\@tempcntb\the\@tempcnta\else
   {\advance\@tempcnta\@ne\ifnum\@tempcnta=\@tempcntb \else \def\@citea{--}\fi
    \advance\@tempcnta\m@ne\the\@tempcnta\@citea\the\@tempcntb}\fi\fi}
 
\makeatother
\begin{titlepage}
\pagenumbering{roman}
\CERNpreprint{\DpPaperGroup}{\DpPaperRef} 
\date{{\small\DpDate}} 
\title{\DpTitle} 
\address{\DpAuthors} 
\begin{shortabs} 
\noindent
%
\noindent
Measurements of on-shell $ZZ$ production are described, using 
data collected by DELPHI in 1997 and 1998, at centre-of-mass energies 
 $\sqrt s$ = 
182.6 GeV and 188.6 GeV respectively. 
Results obtained in each of the final states
$q \bar{q} q \bar{q}$,
$\mu^+ \mu^- q \bar{q}$, 
$e^+e^- q \bar{q}$, 
$\nu \bar{\nu} q \bar{q}$, 
$l^+l^-l^+l^-$, and
$\nu \bar{\nu} l^+l^-$
are presented. 
The measured cross-sections for on-shell $ZZ$ 
production via the tree-level doubly-resonant
graphs (NC02) are:
\begin{tabbing}
~~~~~~~~~~\=$ \sigma_{\tiny NC02}$~(182.6 GeV) 
          \== 0.38 $ \pm $ $0.18 ~(stat)$ $ \pm $ $0.04 ~(syst)$ pb, \\
~~\\
          \>$ \sigma_{\tiny NC02}$~(188.6   GeV) 
          \>= 0.60 $ \pm $ $0.13 ~(stat)$ $ \pm $ $0.07 ~(syst)$ pb.
\end{tabbing}
They are consistent with the Standard Model expectations of
0.25 pb and 0.65 pb at each energy.
\end{shortabs}
\vfill
\begin{center}
\DpSubmit \ \\ 
\DpComment \ \\
\DpEMail \ \\
\end{center}
\vfill
\clearpage
\headsep 10.0pt
\addtolength{\textheight}{10mm}
\addtolength{\footskip}{-5mm}
\begingroup
%
\newcommand{\DpName}[2]{\hbox{#1$^{\ref{#2}}$},\hfill}
\newcommand{\DpNameTwo}[3]{\hbox{#1$^{\ref{#2},\ref{#3}}$},\hfill}
\newcommand{\DpNameThree}[4]{\hbox{#1$^{\ref{#2},\ref{#3},\ref{#4}}$},\hfill}
\newskip\Bigfill \Bigfill = 0pt plus 1000fill
\newcommand{\DpNameLast}[2]{\hbox{#1$^{\ref{#2}}$}\hspace{\Bigfill}}
%
\footnotesize
\noindent
\DpName{P.Abreu}{LIP}
\DpName{W.Adam}{VIENNA}
\DpName{T.Adye}{RAL}
\DpName{P.Adzic}{DEMOKRITOS}
\DpName{I.Ajinenko}{SERPUKHOV}
\DpName{Z.Albrecht}{KARLSRUHE}
\DpName{T.Alderweireld}{AIM}
\DpName{G.D.Alekseev}{JINR}
\DpName{R.Alemany}{CERN}
\DpName{T.Allmendinger}{KARLSRUHE}
% ...
\normalsize
\endgroup
\titlefoot{Department of Physics and Astronomy, Iowa State
     University, Ames IA 50011-3160, USA
    \label{AMES}}
\titlefoot{Physics Department, Univ. Instelling Antwerpen,
     Universiteitsplein 1, B-2610 Antwerpen, Belgium \\
     \indent~~and IIHE, ULB-VUB,
     Pleinlaan 2, B-1050 Brussels, Belgium \\
     \indent~~and Facult\'e des Sciences,
     Univ. de l'Etat Mons, Av. Maistriau 19, B-7000 Mons, Belgium
    \label{AIM}}
\titlefoot{Physics Laboratory, University of Athens, Solonos Str.
     104, GR-10680 Athens, Greece
    \label{ATHENS}}
\titlefoot{Department of Physics, University of Bergen,
     All\'egaten 55, NO-5007 Bergen, Norway
    \label{BERGEN}}
\titlefoot{Dipartimento di Fisica, Universit\`a di Bologna and INFN,
     Via Irnerio 46, IT-40126 Bologna, Italy
    \label{BOLOGNA}}
\titlefoot{Centro Brasileiro de Pesquisas F\'{\i}sicas, rua Xavier Sigaud 150,
     BR-22290 Rio de Janeiro, Brazil \\
     \indent~~and Depto. de F\'{\i}sica, Pont. Univ. Cat\'olica,
     C.P. 38071 BR-22453 Rio de Janeiro, Brazil \\
     \indent~~and Inst. de F\'{\i}sica, Univ. Estadual do Rio de Janeiro,
     rua S\~{a}o Francisco Xavier 524, Rio de Janeiro, Brazil
    \label{BRASIL}}
\titlefoot{Comenius University, Faculty of Mathematics and Physics,
     Mlynska Dolina, SK-84215 Bratislava, Slovakia
    \label{BRATISLAVA}}
\titlefoot{Coll\`ege de France, Lab. de Physique Corpusculaire, IN2P3-CNRS,
     FR-75231 Paris Cedex 05, France
    \label{CDF}}
\titlefoot{CERN, CH-1211 Geneva 23, Switzerland
    \label{CERN}}
\titlefoot{Institut de Recherches Subatomiques, IN2P3 - CNRS/ULP - BP20,
     FR-67037 Strasbourg Cedex, France
    \label{CRN}}
\titlefoot{Now at DESY-Zeuthen, Platanenallee 6, D-15735 Zeuthen, Germany
    \label{DESY}}
\titlefoot{Institute of Nuclear Physics, N.C.S.R. Demokritos,
     P.O. Box 60228, GR-15310 Athens, Greece
    \label{DEMOKRITOS}}
\titlefoot{FZU, Inst. of Phys. of the C.A.S. High Energy Physics Division,
     Na Slovance 2, CZ-180 40, Praha 8, Czech Republic
    \label{FZU}}
\titlefoot{Dipartimento di Fisica, Universit\`a di Genova and INFN,
     Via Dodecaneso 33, IT-16146 Genova, Italy
    \label{GENOVA}}
\titlefoot{Institut des Sciences Nucl\'eaires, IN2P3-CNRS, Universit\'e
     de Grenoble 1, FR-38026 Grenoble Cedex, France
    \label{GRENOBLE}}
\titlefoot{Helsinki Institute of Physics, HIP,
     P.O. Box 9, FI-00014 Helsinki, Finland
    \label{HELSINKI}}
\titlefoot{Joint Institute for Nuclear Research, Dubna, Head Post
     Office, P.O. Box 79, RU-101 000 Moscow, Russian Federation
    \label{JINR}}
\titlefoot{Institut f\"ur Experimentelle Kernphysik,
     Universit\"at Karlsruhe, Postfach 6980, DE-76128 Karlsruhe,
     Germany
    \label{KARLSRUHE}}
\titlefoot{Institute of Nuclear Physics and University of Mining and Metalurgy,
     Ul. Kawiory 26a, PL-30055 Krakow, Poland
    \label{KRAKOW}}
\titlefoot{Universit\'e de Paris-Sud, Lab. de l'Acc\'el\'erateur
     Lin\'eaire, IN2P3-CNRS, B\^{a}t. 200, FR-91405 Orsay Cedex, France
    \label{LAL}}
\titlefoot{School of Physics and Chemistry, University of Lancaster,
     Lancaster LA1 4YB, UK
    \label{LANCASTER}}
\titlefoot{LIP, IST, FCUL - Av. Elias Garcia, 14-$1^{o}$,
     PT-1000 Lisboa Codex, Portugal
    \label{LIP}}
\titlefoot{Department of Physics, University of Liverpool, P.O.
     Box 147, Liverpool L69 3BX, UK
    \label{LIVERPOOL}}
\titlefoot{LPNHE, IN2P3-CNRS, Univ.~Paris VI et VII, Tour 33 (RdC),
     4 place Jussieu, FR-75252 Paris Cedex 05, France
    \label{LPNHE}}
\titlefoot{Department of Physics, University of Lund,
     S\"olvegatan 14, SE-223 63 Lund, Sweden
    \label{LUND}}
\titlefoot{Universit\'e Claude Bernard de Lyon, IPNL, IN2P3-CNRS,
     FR-69622 Villeurbanne Cedex, France
    \label{LYON}}
\titlefoot{Univ. d'Aix - Marseille II - CPP, IN2P3-CNRS,
     FR-13288 Marseille Cedex 09, France
    \label{MARSEILLE}}
\titlefoot{Dipartimento di Fisica, Universit\`a di Milano and INFN-MILANO,
     Via Celoria 16, IT-20133 Milan, Italy
    \label{MILANO}}
\titlefoot{Dipartimento di Fisica, Univ. di Milano-Bicocca and
     INFN-MILANO, Piazza delle Scienze 2, IT-20126 Milan, Italy
    \label{MILANO2}}
\titlefoot{Niels Bohr Institute, Blegdamsvej 17,
     DK-2100 Copenhagen {\O}, Denmark
    \label{NBI}}
\titlefoot{IPNP of MFF, Charles Univ., Areal MFF,
     V Holesovickach 2, CZ-180 00, Praha 8, Czech Republic
    \label{NC}}
\titlefoot{NIKHEF, Postbus 41882, NL-1009 DB
     Amsterdam, The Netherlands
    \label{NIKHEF}}
\titlefoot{National Technical University, Physics Department,
     Zografou Campus, GR-15773 Athens, Greece
    \label{NTU-ATHENS}}
\titlefoot{Physics Department, University of Oslo, Blindern,
     NO-1000 Oslo 3, Norway
    \label{OSLO}}
\titlefoot{Dpto. Fisica, Univ. Oviedo, Avda. Calvo Sotelo
     s/n, ES-33007 Oviedo, Spain
    \label{OVIEDO}}
\titlefoot{Department of Physics, University of Oxford,
     Keble Road, Oxford OX1 3RH, UK
    \label{OXFORD}}
\titlefoot{Dipartimento di Fisica, Universit\`a di Padova and
     INFN, Via Marzolo 8, IT-35131 Padua, Italy
    \label{PADOVA}}
\titlefoot{Rutherford Appleton Laboratory, Chilton, Didcot
     OX11 OQX, UK
    \label{RAL}}
\titlefoot{Dipartimento di Fisica, Universit\`a di Roma II and
     INFN, Tor Vergata, IT-00173 Rome, Italy
    \label{ROMA2}}
\titlefoot{Dipartimento di Fisica, Universit\`a di Roma III and
     INFN, Via della Vasca Navale 84, IT-00146 Rome, Italy
    \label{ROMA3}}
\titlefoot{DAPNIA/Service de Physique des Particules,
     CEA-Saclay, FR-91191 Gif-sur-Yvette Cedex, France
    \label{SACLAY}}
\titlefoot{Instituto de Fisica de Cantabria (CSIC-UC), Avda.
     los Castros s/n, ES-39006 Santander, Spain
    \label{SANTANDER}}
\titlefoot{Dipartimento di Fisica, Universit\`a degli Studi di Roma
     La Sapienza, Piazzale Aldo Moro 2, IT-00185 Rome, Italy
    \label{SAPIENZA}}
\titlefoot{Inst. for High Energy Physics, Serpukov
     P.O. Box 35, Protvino, (Moscow Region), Russian Federation
    \label{SERPUKHOV}}
\titlefoot{J. Stefan Institute, Jamova 39, SI-1000 Ljubljana, Slovenia
     and Laboratory for Astroparticle Physics,\\
     \indent~~Nova Gorica Polytechnic, Kostanjeviska 16a, SI-5000 Nova Gorica, Slovenia, \\
     \indent~~and Department of Physics, University of Ljubljana,
     SI-1000 Ljubljana, Slovenia
    \label{SLOVENIJA}}
\titlefoot{Fysikum, Stockholm University,
     Box 6730, SE-113 85 Stockholm, Sweden
    \label{STOCKHOLM}}
\titlefoot{Dipartimento di Fisica Sperimentale, Universit\`a di
     Torino and INFN, Via P. Giuria 1, IT-10125 Turin, Italy
    \label{TORINO}}
\titlefoot{Dipartimento di Fisica, Universit\`a di Trieste and
     INFN, Via A. Valerio 2, IT-34127 Trieste, Italy \\
     \indent~~and Istituto di Fisica, Universit\`a di Udine,
     IT-33100 Udine, Italy
    \label{TU}}
\titlefoot{Univ. Federal do Rio de Janeiro, C.P. 68528
     Cidade Univ., Ilha do Fund\~ao
     BR-21945-970 Rio de Janeiro, Brazil
    \label{UFRJ}}
\titlefoot{Department of Radiation Sciences, University of
     Uppsala, P.O. Box 535, SE-751 21 Uppsala, Sweden
    \label{UPPSALA}}
\titlefoot{IFIC, Valencia-CSIC, and D.F.A.M.N., U. de Valencia,
     Avda. Dr. Moliner 50, ES-46100 Burjassot (Valencia), Spain
    \label{VALENCIA}}
\titlefoot{Institut f\"ur Hochenergiephysik, \"Osterr. Akad.
     d. Wissensch., Nikolsdorfergasse 18, AT-1050 Vienna, Austria
    \label{VIENNA}}
\titlefoot{Inst. Nuclear Studies and University of Warsaw, Ul.
     Hoza 69, PL-00681 Warsaw, Poland
    \label{WARSZAWA}}
\titlefoot{Fachbereich Physik, University of Wuppertal, Postfach
     100 127, DE-42097 Wuppertal, Germany
    \label{WUPPERTAL}}
\addtolength{\textheight}{-10mm}
\addtolength{\footskip}{5mm}
\clearpage
\headsep 30.0pt
\end{titlepage}
%
\pagenumbering{arabic} 
\setcounter{footnote}{0} %
\large

%

\section{Introduction}

The study of doubly resonant production of $Z$ bosons is a relatively new 
topic. The first evidence~\cite{vanc} was accumulated 
during 1997, when LEP was operated at a centre-of-mass energy of 
182.6 GeV, corresponding to the threshold for this channel.
In this letter we present
measurements of the production cross-section both from that run and from the
1998 run at 188.6 GeV.

There are several motivations for studying this channel. Firstly, it is 
necessary to check that the observed production rate of any expected
physical process seen for the first time in a new energy domain is well 
accounted for by the Standard Model. Deviations from predictions
could be interpreted as a signal for new physics beyond the Standard Model,
manifesting itself through anomalous production~\cite{hdoublet,grav}, 
for instance by means of anomalous neutral-current triple gauge boson
couplings~\cite{hagiwara}. Secondly, $ZZ$ production forms 
an irreducible background to the Higgs search at LEP when the mass of the 
Higgs boson is close to that of the $Z$~\cite{higgspre}. In this context 
the results obtained also give some indication of the reliability of the 
techniques used in the Higgs search.


In what follows, the data sets and simulations used are described and the 
signal definition which was adopted is discussed. The event selections 
developed in the six sub-channels which were analysed are then presented. 
Results are given in the form of a comparison of the numbers of found and 
predicted selected events, together with an evaluation of the main systematic 
effects. Finally, the combination of the sub-channel results into overall $ZZ$ 
cross-sections is described and the overall measurements are compared to the 
Standard Model expectation.

\section{Data samples}

DELPHI took data at centre-of-mass energies of 182.6 GeV in 1997 and 
188.6 GeV in 1998, with integrated luminosities of 54 pb$^{-1}$ and
158 pb$^{-1}$ respectively. A detailed description of the detector and 
a review of its performance can be found in~\cite{DELPHIDET,DELPHIPER}.
The detector was not changed in recent years, except for upgrades 
of the vertex detector~\cite{VD}, and the addition of a set of scintillator
counters to veto photons in blind regions of the electromagnetic 
calorimetry, at polar angles near 40$^\circ$~ and 90$^\circ$~.

Simulated events were produced with the DELPHI simulation program 
{\tt DELSIM}\cite{DELPHIPER} and were then passed through the same
reconstruction and analysis chain as the data. 
The generation of processes leading to four-fermion final 
states was done with {\tt EXCALIBUR}\cite{EXCALIBUR}, relying on 
{\tt JETSET} 7.4 \cite{JETSET} for quark fragmentation. 
{\tt GRC4F}\cite{GRC4F} was used as a complementary generator for 
four-fermion final states resulting
from $W e \nu_e$ processes when $\cos \theta_e$   $>$ 0.9999. 
Two fermion processes $e^+ e^- \rightarrow\ f \bar{f} (+ n\gamma)$
were generated using {\tt PYTHIA} \cite{JETSET}, 
$e^+ e^- \rightarrow\ \mu^+ \mu^- (+ n\gamma)$ and
$e^+ e^- \rightarrow\ \tau^+ \tau^- (+ n\gamma)$ with 
{\tt KORALZ}\cite{KORALZ},
and $e^+ e^- \rightarrow\ e^+ e^- (+ n\gamma)$ with 
{\tt BHWIDE}\cite{BHWIDE}. Two-photon interactions were generated 
using {\tt TWOGAM}~\cite{TWOGAM} and {\tt BDK}~\cite{BDK}.




\section{Signal definition}

\begin{figure}[tbh]
  \begin{center}
\mbox{\epsfysize=9.0cm\epsfxsize=13cm\epsffile{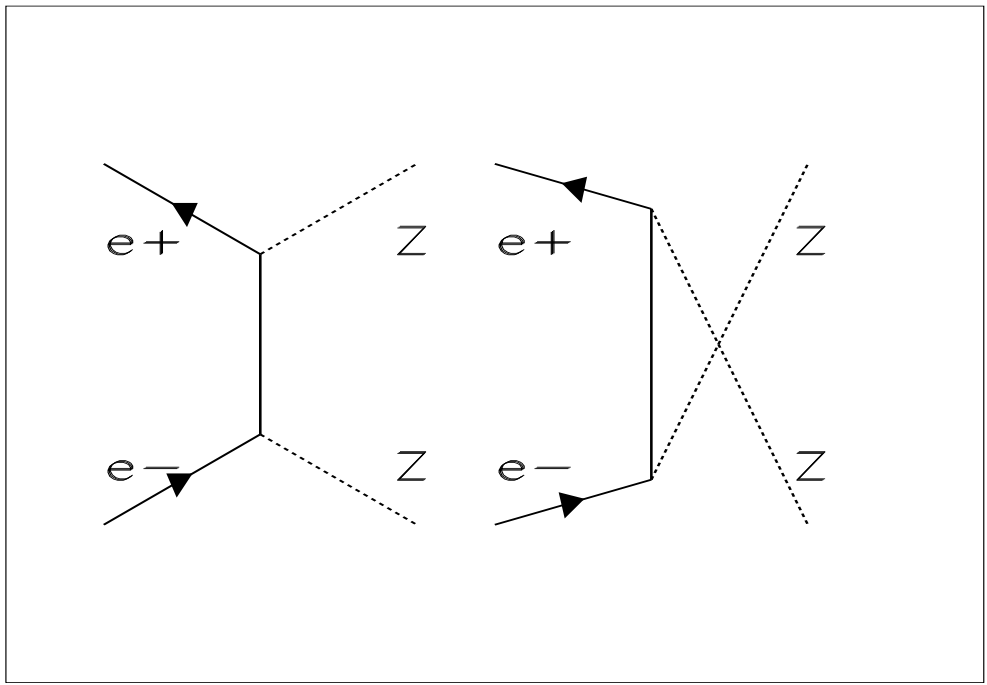}}
    \caption{ The Feynman graphs for on-shell $ZZ$ production 
(referred to as the NC02 graphs)
      }
    \label{fig:fnc02}
  \end{center}
\end{figure}

The region of phase-space at high di-fermion masses must be isolated
to measure the $ZZ$ production cross-section. In order to interpret 
the measurement in terms of the tree-level doubly-resonant graphs shown 
in Figure~\ref{fig:fnc02} (referred to as the NC02 graphs) 
the presence of other 
four-fermion processes in this region must be taken into account.
For this purpose the signal was defined in the simulation by 
requiring that the 
generated masses of the two appropriate pairings of final-state fermions 
be within 10 GeV/c$^2$ of the nominal $Z$ mass. 
This choice of mass window maximized
the sensitivity to the NC02 graphs while minimizing contributions from other 
four-fermion processes. Events with the correct flavour composition but 
which fell outside this generator-level mass window were 
considered as background. A scaling factor R,

\begin{equation}
{\rm R} = {\sigma_{\rm NC02}^{\rm total} \over \sigma_{\rm 4f}^{\rm window}},
\end{equation}
 
\noindent 
was then calculated at generator level
to enable conversion of the measured 
total four-fermion cross-section within the mass window, 
$\sigma_{\rm 4f}^{\rm window}$, into the total NC02 cross-section, 
$\sigma_{\rm NC02}^{\rm total}$.
The scaling factors obtained for each channel 
are shown in table~\ref{table:scalingfac}. 
Most of the values are close to unity, confirming that the defined 
region is dominated by on-shell $ZZ$ production. 
Several sources of bias which could result from this procedure were 
investigated. The most relevant among these were:
\begin{itemize}
\item
In the channel $e^+ e^- q \bar{q}$, the scaling factors are smaller 
than unity. This arises because of significant contributions 
from the single-resonant process $e^+ e^- Z$. Since the electrons from
this process tend to be peaked in the forward directions and since 
the efficiencies to identify electrons are reduced in these regions, there is
a bias for this particular channel from taking a scaling factor 
averaged over the full solid angle. The magnitude of this
bias was estimated by computing the expected cross-sections 
using this procedure in the barrel and forward regions separately. A 2.75\% 
correction was derived and incorporated into the scaling factor quoted in 
table~\ref{table:scalingfac}.
\item
In the channel $l^+l^-l^+l^-$ the factor is significantly smaller than 
unity. This arises because of the intrinsic ambiguity in pairing, existing 
in cases with four leptons of the same family, for which  
there are significant contributions 
in the signal window from $Z \gamma^{*}$ processes. 
As an additional
cross-check for this channel the measurement was also repeated using 
an additive correction procedure (see the corresponding section),
yielding fully consistent results.
\end{itemize}

\begin{table}[htb]
\begin{center}
\begin{tabular}
{|c|c|c|}
\hline

Final State                    & 183 GeV      & 189 GeV       \\ 
\hline
$q \bar{q} q \bar{q}$          & 1.26         & 1.07          \\ 
$\nu \bar{\nu} q \bar{q}$      & 1.14         & 1.05          \\ 
$\mu^+ \mu^- q \bar{q}$        & 1.19         & 1.05          \\ 
$e^+e^- q \bar{q}$             & 0.83         & 0.88          \\ 
$l^+l^-l^+l^-$                 & 0.46         & 0.59          \\ 
$\nu \bar{\nu} l^+l^-$         & 1.00         & 1.00          \\ 
\hline

\end{tabular}
\end{center}
\caption[.]{
\label{table:scalingfac}
Scaling factors R computed at generator level
to convert the measured total four-fermion cross-section 
within the signal defining mass window $|M_{f{\bar f}} - M_Z| < 10$ GeV/c$^2$, 
$\sigma_{\rm 4f}^{\rm window}$, into the total NC02 cross-section, 
$\sigma_{\rm NC02}^{\rm total}$.}
\end{table}

%
%
%
\section{Four jets}
%
%
%

The \zz $\rightarrow$ \qqqq process represents 49\% of the \zz final 
states and results typically in events with four or more jets. The principal
backgrounds arise from \ww and \qcd processes, which can lead to similar
multijet topologies. The main ingredients used to isolate the signal 
were the $b$-tagging of the jets, topological information quantifying 
their separation, and reconstructed di-jet masses. A pre-selection was 
first applied to isolate hadronic events with at least four reconstructed 
jets, as in \cite{ww183}. Four and five jet events were treated separately
throughout the analysis\footnote{Events with more than 
five reconstructed jets were forced into a five-jet configuration.}. 

In order to optimally use the information from the $b$-tagging, 
the ratio of the distributions of the combined $b$-tagging 
variable \cite{btag} 
for $b$ and non-$b$ jets were parametrized in three different angular regions,
using simulated \z decays. Using these parametrisations and the measured value 
of the combined $b$-tagging variable for each jet, the probability that it 
originated from the fragmentation of a $b$ or a non-$b$ quark was evaluated. 

In order to fully exploit the information from the mass reconstruction, all
possible di-jet pairings were used, following the method outlined 
in~\cite{ww183}. For each of the physics hypotheses \zz, \ww and \qcd, 
the expected probability distributions for the different mass combinations 
in an event were expressed in analytical form, by means of products of 
Breit-Wigners and phase-space factors for the correctly paired mass 
combinations in \ww and \zz events, and of flat spectra for all the 
other cases.
Using the event-by-event jet errors, measured $\chi^2$ probability 
distributions were obtained, as functions of the possible di-jet masses, 
by kinematic fits~\cite{kinfit} requiring four-momentum conservation and 
equality of each di-jet mass with the tested value. These two-dimensional 
measurement probabilities were then convoluted with the expected distributions 
to quantify the compatibility of each pairing with the \zz\, \ww and \qcd 
hypotheses from the mass information alone.

Finally the topological variable ${\rm E_{min} \cdot \alpha_{min}}$ was 
defined, where  ${\rm E_{min}}$ is the minimum jet-energy and 
${\rm \alpha_{min}}$ the minimum opening angle between any jet pair. 
The ratio of two-fermion and four-fermion events was parametrized 
as a function of this variable using simulated data for four- and 
five-jet events separately.

\begin{figure}[tbh]
\begin{center}
    \mbox{\epsfysize=17.0cm\epsffile{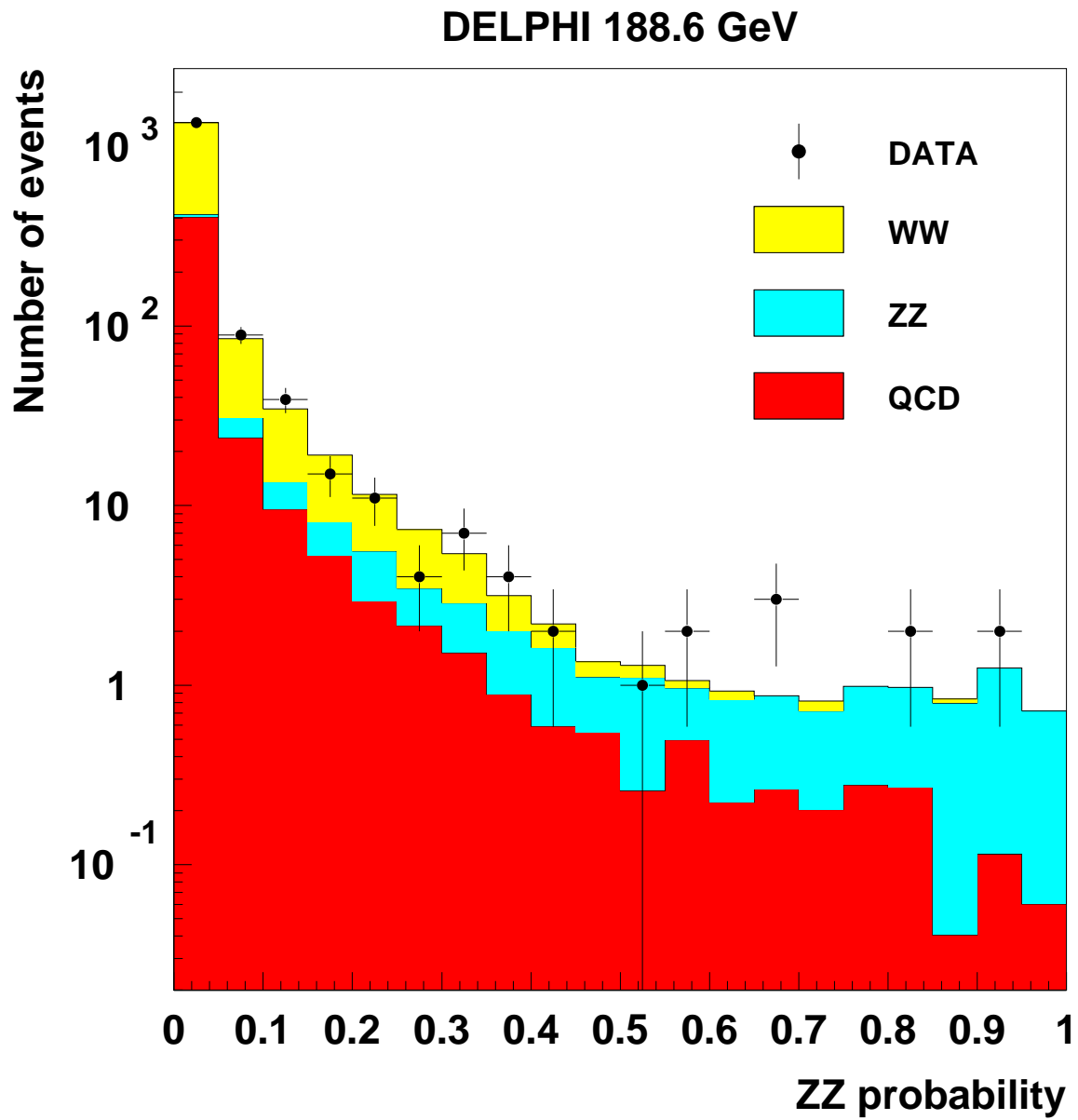}}
\end{center}
\caption{Distribution of the \zz probability for all events at 188.6 GeV in
the four-jet channel}
\label{fig:pzz189}
\end{figure}

Looking for all possible hadronic \zz final states and using the 
predicted SM cross-sections and branching ratios into the different quark 
configurations, the $b$-like probability per jet, the topological 
information per event and the mass information per pairing, a combined 
variable quantifying the compatibility of any event with the \zz hypothesis 
was constructed. This combined variable was shown to behave as a genuine 
probability for the \zz hypothesis, and provided high discriminating power. 
The distribution of this variable at 188.6 GeV is shown in 
Figure~\ref{fig:pzz189}. Results obtained by cutting on this 
variable are presented in table~\ref{tab:presel189}.

\begin{table}[ht]
\begin{center}
 \begin{tabular}{ | c | r | r | r | r | r | r | }
   \hline
    {\bf 182.6 GeV}                & data & Total MC~ & \zz   & \zg  & \ww  & \qcd \\
   \hline
   After pre-selection             & \hspace{0.20cm}535 & \hspace{0.20cm}510.8 & \hspace{0.20cm}6.73  &  \hspace{0.20cm}7.06 & \hspace{0.20cm}336.2 & 160.8 \\
   At maximum $\epsilon \cdot$p   & 2 &  1.68  & 1.01  & 0.10 & 0.11  & 0.46 \\
   \hline
 \end{tabular}

\vspace{0.5cm}

 \begin{tabular}{ | c | r | r | r | r | r | r | }
   \hline
    {\bf 188.6 GeV}                & data & Total MC~ & \zz   & \zg  & \ww    & \qcd \\
   \hline
   After pre-selection             & 1538 & 1536.0 & 41.8  & 17.7 & 1039.6 & 437.0 \\
   At maximum $\epsilon \cdot $p   & 42   & 43.06  & 15.54 & 1.41 & 16.01  & 10.10 \\
   At S/B = 2.92                   & 7    & 7.69   & 5.73  & 0.38 &  0.28  & 1.30 \\
   \hline
 \end{tabular}
\caption{The observed and expected number of selected events 
after the pre-selection at 182.6 GeV and at 188.6 GeV, for 
integrated luminosities of 53 pb$^{-1}$ and  158 pb$^{-1}$ respectively. 
The two last lines give 
the number of events left at the maximum of efficiency $\times$ 
purity ($\epsilon \cdot $p) and at a signal-to-background ratio, S/B, 
of about 3.}
\label{tab:presel189}
\end{center}
\end{table}

Systematic effects resulting from uncertainties in signal efficiency, in the 
conversion factor to translate the results to a NC02 cross-section and in 
the predicted background level were studied. The largest errors came from
uncertainties in the modeling of multijet \qcd\ processes with $b$ 
quarks composing the main remaining background, and from the 
sensitivity of the $b$-tagging procedure to the jet multiplicity.
The dominant \qcd\ modeling uncertainty came from the limited precision
with which the $g \rightarrow b \bar {b}$ rate is known, and to a lesser 
extent from uncertainties in the amount of reduction in gluon radiation 
off $b$ quarks (also known as the dead-cone effect). These uncertainties, 
as well as biases in the treatments provided by the generator, have been 
studied~\cite{qcdb} by comparing results of analytic calculations with 
dedicated measurements at LEP-1 and with predictions of the generators 
used in the simulation. Uncertainties in selection efficiencies related 
to the $b$-tagging procedure were studied by comparing efficiencies for 
two and four jet events at LEP-1 energies. By propagating effects from 
these sources into the analysis, a combined systematic uncertainty of 
around 6\% was estimated relative to the expected cross-section.


%
%
%
\section {Jets and a pair of isolated leptons \label{sub:QQLL}}
%
%
%

The decay modes $e^+ e^- \rightarrow \mu^+ \mu^- q \bar{q}$ 
and $e^+ e^- \rightarrow e^+ e^- q \bar{q}$ represent
9\% of the $ZZ$~final states. Events with $\tau^+ \tau^-$ 
pairs were not considered. The two final state leptons are typically 
well isolated from all other particles. This can be used to select such 
events with high efficiency in both the muon and electron channels. 
Events were selected initially without explicit cuts on the masses of 
the final state fermion pairs in order to analyse simultaneously $ZZ$,  
$Z\gamma^*$ events and contributions from other possible diagrams 
leading to final states such as $Ze^+e^-$ or t-channel $\gamma^*$ 
exchange with $Z/\gamma$-strahlung. Mass cuts were then applied to 
isolate the $ZZ$ contribution.

A loose hadronic pre-selection was first applied, requiring that events 
have at least 7 charged particles and a charged energy above 0.30 $\sqrt{s}$. 
To suppress the radiative return to the $Z$ boson, events were rejected 
if a photon with energy more than 60 GeV was found. The selection procedures 
then proceeded in a closely similar way for both $\mu^+ \mu^- q \bar {q}$ and 
$e^+ e^- q \bar {q}$ channels. In order to maximize the lepton
identification efficiency, any charged particle with a momentum 
exceeding 5 GeV/c was considered as a possible lepton candidate 
around which nearby photons, if present, could be clustered. 
This was found to be necessary 
to improve the energy evaluation in the presence of final state radiation, 
and, in the case of electrons, bremsstrahlung. In the case of the 
$e^+ e^- q \bar {q}$ channel, photons with energy between 20~GeV and 60~GeV
were also considered as electron candidates, to recover events in which the 
electron track was not reconstructed.

Events with at least two lepton candidates of the same flavour, opposite
charge and invariant mass exceeding 2 GeV/c$^2$ were then selected. 
All particles except the lepton candidates were clustered into jets 
and a kinematic fit requiring four-momentum conservation was applied,
correcting appropriately the errors on lepton energies in cases where
photons had been added by the clustering procedure.

At least one of the two lepton candidates was required to satisfy 
strong lepton identification criteria, while softer requirements
were specified for the second. Muons were considered as strongly 
identified if selected by the standard DELPHI muon identification 
package \cite{DELPHIPER}, based mainly on finding associated hits 
in the muon chambers. For soft muon identification only a set of 
kinematic and calorimetric criteria were used. Electrons were 
considered as strongly identified when the energy deposited in the 
electromagnetic calorimeter exceeded 60\% of the cluster energy or 
15 GeV and when the energy deposited in the hadron calorimeter was 
reasonably limited. For soft electron identification only requirements 
on the momentum of the charged particle in the cluster and on the
energy deposited in the hadron calorimeter were used. Moreover electron 
candidates originating from applying the clustering procedure around 
a photon were considered as softly identified.

\begin{figure}[tbh]
  \begin{center}
    \mbox{\epsfysize=17.0cm\epsffile{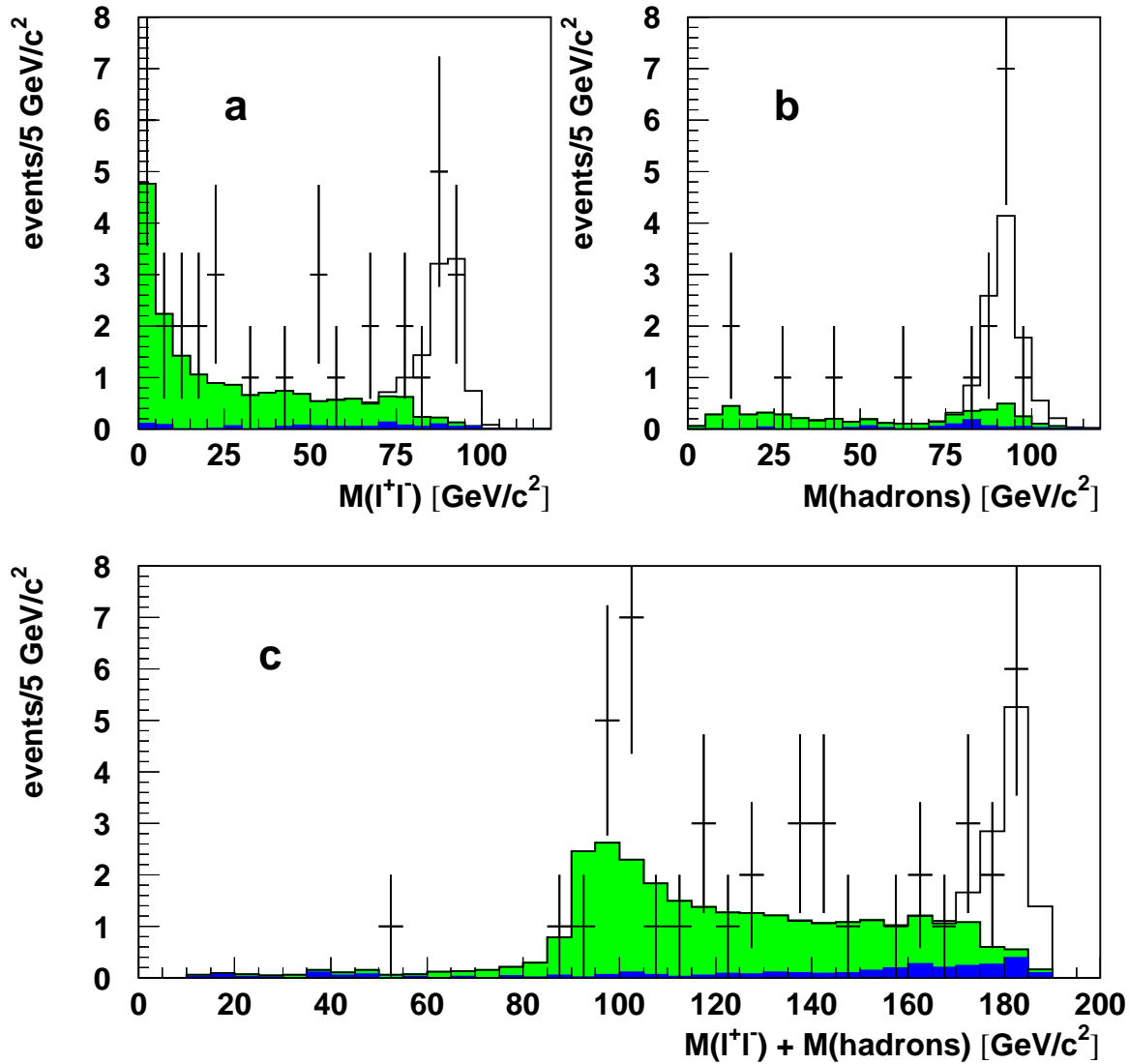}}
    \caption{ a) the distribution of the mass of the $l^+ l^-$ pair when 
      the mass of the hadron system is within 15 GeV/c$^2$
      of the nominal $Z$ mass; b) the distribution of the mass of 
      the hadron system when 
      the mass of the $l^+ l^-$ pair is within 15 GeV/c$^2$
      of the nominal $Z$ mass;
      c) the distribution of the sum of the masses of the $l^+ l^-$ pair 
      and of the hadron system.
      The points are the data taken at 182.6 GeV and at 188.6 GeV, 
      and the histogram is the simulation prediction.
      The contribution from the signal, as defined in section
      3, is indicated by the empty histogram. The contributions from 
      backgrounds are indicated with light grey when arising from 
      from $l^+l^-q{\bar q}$ final states, and with dark filling
      otherwise.
      }
    \label{llqq:fig4}
  \end{center}
\end{figure}

Two discriminating variables were then defined for final event selection: 
$P_t^{min}$, the lesser of the transverse momenta of the lepton candidates 
with respect to their nearest jet and the $\chi^2$ per degree of freedom 
of the kinematic fit. The distribution of the mass of one fermion pair 
($l^+ l^-$ or $q \bar{q}$) when the mass of the second pair is within 
15 GeV/c$^2$ of the nominal $Z$ mass is shown in figure \ref{llqq:fig4}a,b. 
The distribution of the sum of masses of two fermion pairs is shown in
figure \ref{llqq:fig4}c. The observed distributions are in reasonable
agreement with the predictions from simulation. To select on-shell $ZZ$ 
production, cuts were placed simultaneously on the masses of the $l^+ l^-$ pair, 
on the remaining hadron system, and on their sum, taking into account in the 
performance optimization the different mass resolutions of these final states 
and the presence of the single-resonant process, $e^+ e^- Z$ in the case of the
$e^+e^- q \bar{q}$ channel. The observed and predicted numbers of selected 
events are shown in table~\ref{llqq:tab3}. The background is divided into 
two parts: the first one comes from $l^+l^- q \bar{q}$ events outside the 
generation-level signal window, and the second one comes from other processes, 
principally $W^+W^-$, other $ZZ$ decays and 
(in the case of $e^+e^- q \bar{q}$) 
$q \bar{q} (+ \gamma)$ production. For the $\mu^+ \mu^- q \bar {q}$ channel, 
the efficiencies of the selection were 0.89 $\pm$ 0.02 and 0.86 
$\pm$ 0.01 at 182.6 and 188.6 GeV respectively. 
For the $e^+ e^- q \bar {q}$ channel, they 
were 0.73 $\pm$ 0.03 and 0.72 $\pm$ 0.02 respectively.
 
\begin{table}[tbh]
  \begin{center}
    \begin{tabular}{|c|c|c|c|c|c|c|}
      \hline
$E_{cms}$&channel&data&Total MC&MC signal& MC $l^+l^-q \bar{q}$ &MC other \\
GeV      &       &    &       &         & backgr.              & backgr. \\
      \hline
182.6 & $\mu^+\mu^- q \bar{q}$ & 3  & $0.52\pm0.03$ & $0.48\pm0.03$  &
                                      $0.04\pm0.01$ & $0.00\pm0.01$ \\
      & $e^+e^- q \bar{q}$     & 0  & $0.69\pm0.07$ & $0.55\pm0.05$  &
                                      $0.10\pm0.02$ & $0.04\pm0.04$ \\
                                                             \cline{2-7}
      & Total                  & 3  & $1.21\pm0.08$ & $1.03\pm0.06$  &
                                      $0.14\pm0.02$ & $0.04\pm0.04$ \\
                                                             \hline
                                                             \hline
188.6 & $\mu^+\mu^- q \bar{q}$ & 5  & $4.15\pm0.11$ & $3.92\pm0.10$  &
                                      $0.19\pm0.02$ & $0.04\pm0.03$ \\
      & $e^+e^- q \bar{q}$     & 3  & $4.04\pm0.15$ & $3.55\pm0.13$  &
                                      $0.35\pm0.04$ & $0.14\pm0.06$ \\
                                                             \cline{2-7}
      & Total                  & 8  & $8.19\pm0.19$ & $7.47\pm0.16$  &
                                      $0.54\pm0.04$ & $0.18\pm0.07$ \\
                                                             \hline
                                                             \hline
All   & $\mu^+\mu^- q \bar{q}$ & 8  & $4.67\pm0.11$ & $4.40\pm0.11$  &
                                      $0.23\pm0.02$ & $0.04\pm0.03$ \\
      & $e^+e^- q \bar{q}$     & 3  & $4.73\pm0.17$ & $4.10\pm0.14$  &
                                      $0.45\pm0.05$ & $0.18\pm0.07$ \\
                                                             \cline{2-7}
      & Total                  & 11 & $9.40\pm0.20$ & $8.50\pm0.18$  &
                                      $0.68\pm0.05$ & $0.22\pm0.08$ \\
                                                             \hline
    \end{tabular}
    \caption []{ The observed and expected number of selected
      $l^+ l^- q \bar{q}$ events - candidates to the on-shell $ZZ$
      production. The errors shown are due to the simulation statistics only.
      The expected numbers of the signal events are for the SM cross-section
      of the on-shell $ZZ$ production.}
    \label{llqq:tab3}
  \end{center}
\end{table}

Several sources of systematic errors were investigated. Uncertainties in the 
lepton identification were estimated comparing semileptonic $W^+W^-$ events 
selected in data and simulation using the strong lepton identification 
criteria.
Uncertainties in signal efficiencies from the description of the kinematic 
observables used were evaluated comparing the $P_t$ and $\chi^2$ distributions 
in data and simulation for all $l^+l^- q \bar{q}$ events selected without mass cuts. 
Corresponding uncertainties in background levels were evaluated by comparing 
samples of events selected in data and in simulation requiring both isolated 
tracks not to be identified as leptons, while maintaining all other criteria.
Finally, uncertainties in the background level in the $e^+e^- q \bar{q}$ 
channel 
from fake electrons were studied with $Z\gamma$ events selected in data and in 
simulation with purely kinematic criteria. Propagating the found differences 
to the 
final stage of the analysis yielded a combined systematic error on the 
efficiency 
to select $l^+l^- q \bar{q}$ events of $\pm$3.0\%. The uncertainty 
resulting for 
the background level amounted to about $\pm$15\%.

\section{Jets and missing energy}
%
%
%

The decay mode $\nu \bar{\nu} q \bar{q}$ represents 28\% of the $ZZ$~final 
states. Its signature is a pair of jets relatively acoplanar with the beam 
and with visible and recoil masses compatible with the $Z$ mass. The most 
difficult backgrounds arise from single resonant $W e \nu_e$ processes, 
from $WW$ processes where one of the $W$ decays into $\tau \nu_{\tau}$, 
and from $q\bar{q}$ events, accompanied or not by isolated photons escaping 
detection, in which one or two of the jets were badly reconstructed.

A pre-selection was first applied to remove the bulk of the background. 
Multihadronic annihilation events were selected by requiring that the number 
of charged particles be larger than 8, that the track of at least one 
charged particle, with a transverse momentum larger than 1.5~GeV/c, 
extrapolate back to within 200~$\mu$m of the primary vertex in the plane 
transverse to the beam axis, that the total charged energy of the event 
exceed 10\% of the centre-of-mass energy and that its raw visible mass be 
in the range 81~$\pm$~22~GeV/c$^2$. To reject radiative returns to the 
$Z$ with energetic photons emitted in blind regions of the electromagnetic 
calorimetry (at polar angles near 40$^\circ$~ and 90$^\circ$~) signals from 
dedicated scintillator counters were used.

A combined discriminant variable was then constructed using the Iterative 
Discriminant Analysis program (IDA)~\cite{IDA} to calculate a second order 
polynomial from seven event variables, selected based on their discriminating 
power and independence:

\begin{itemize}
\item The minimum polar angle defining a cone in the positive 
and negative beam directions containing 15\% of the total visible energy. 
\item The logarithm of the acoplanarity scaled by the sine of the minimum 
polar angle between a jet direction and the beam axis.
\item The total reconstructed energy.
\item The maximum transverse momentum between any particle and a jet. 
\item The energy of the more energetic among the reconstructed jets.
\item The thrust, computed in the rest frame of the visible system. 
\item The acollinearity of the jets.
\end{itemize}

Moreover, to concentrate on the signal region when optimizing the
second-order discriminant function, very loose cuts were applied on these
variables, at values corresponding to the tails of the signal distributions 
and removing each about one percent of it. Finally, the total energy of 
hypothetical photons escaping in the beam direction was estimated from 
the two jet directions and was required to be less than 85\% of 
the value expected for radiative returns to the $Z$.

\begin{figure}[tbh]
  \begin{center}
{\Large \leftline {a)}}
   \mbox{\epsfxsize=11.0cm\epsfysize=9.0cm\epsffile{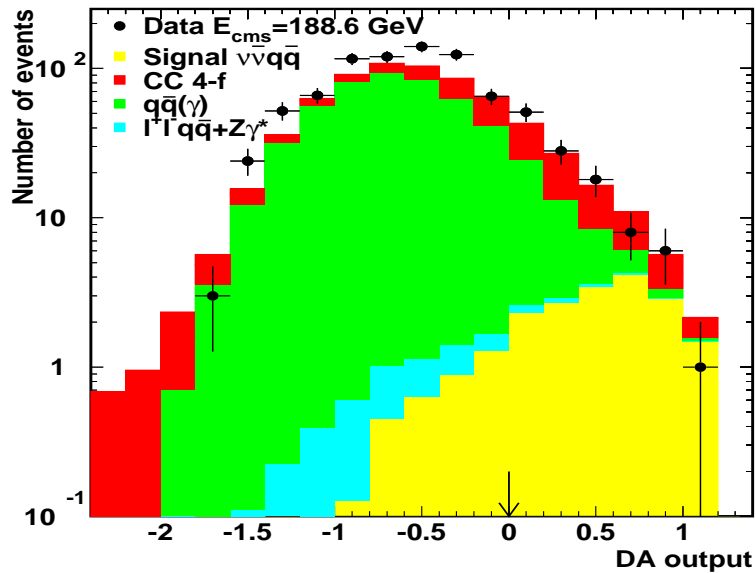}}    
{\Large \leftline {b)}}
   \mbox{\epsfxsize=11.0cm\epsfysize=9.0cm\epsffile{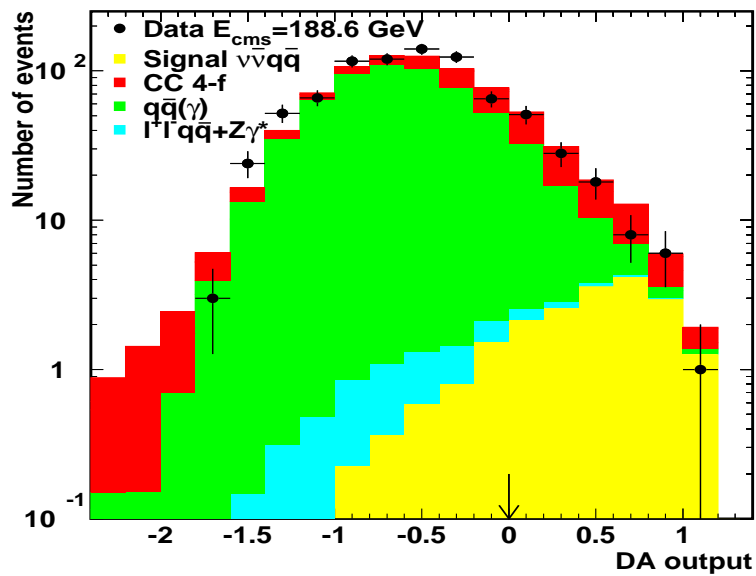}}  
  \caption{ $\nu \bar{\nu} q \bar{q}$ channel: The second-order 
           discriminant for the data at 188.6~GeV, for the expected 
           signal (light) and for the backgrounds from
           $l^+ l^- q \bar{q}$ and $Z\gamma^*$ (light grey),
           $q\bar{q}(\gamma$) (dark grey) and 
           $W e \nu_e$ and $WW$ (dark, CC 4-f in the legend) 
           processes are shown,
           using the nominal (plot a) and corrected (plot b) 
           simulation, following the procedure described in the text.
           The main effect from the correction procedure
           was to increase the $q\bar{q}(\gamma$) background component.
           The arrow indicates the cut used to remove the part of the
           distribution where the $q\bar{q}(\gamma)$ background dominates, 
           in order to reduce the systematic error as described in the text. 
           \label{fig:vvqq_da_189_sys}}
  \end{center}
\end{figure}

The comparison of selected data and simulation rates for the signal 
and background components is shown in Figure~\ref{fig:vvqq_da_189_sys}a
(for the 188.6~GeV data) and in table~\ref{tab:vvqqsel}. 
The $ZZ$ 
cross-section at each energy was obtained from a binned maximum likelihood 
fit of this discriminant output with the $ZZ$ signal contribution as 
the only free parameter.

\begin{table}
  \begin{center}
{
    \begin{tabular}{|c|c|c|c|c|c|c|c|}
      \hline
183 GeV        & Data  & Total MC        & Signal       & CC 4-f        & $q\bar{q}\gamma$   & $q\bar{q}l^+l^-$ &  $\gamma\gamma$ \\
      \hline
After pre-selection& 2485  & 2286$\pm$10     & 3.5$\pm$0.1  &  134$\pm$1    & 2005$\pm$4         &  6.9$\pm$0.2     & 137$\pm$9\\
After tail cuts    &  235  &  209$\pm$2      & 3.0$\pm$0.1  &   37$\pm$0.6  &  163$\pm$1         &  1.0$\pm$0.1     & 5.1$\pm$1.6 \\
After final DA cut &   22  & 21.9$\pm$0.45   & 2.4$\pm$0.1  &  8.8$\pm$0.3  & 10.6$\pm$0.4       &  0.07$\pm$0.03     &   0 \\
\hline
      \hline
189 GeV        & Data  & Total MC        & Signal        & CC 4-f       & $q\bar{q}\gamma$ & $q\bar{q}l^+l^-$ &  $\gamma\gamma$ \\
      \hline
After pre-selection& 6723  & 6206$\pm$20     & 21.5$\pm$0.3  &  344$\pm$2   & 5344$\pm$7       & 17.1$\pm$0.5     & 479$\pm$18  \\
After tail cuts    & 824   &  710$\pm$6      & 20.4$\pm$0.3  &  158$\pm$1   &  500$\pm$3       &  3.3$\pm$0.1     & 28.5$\pm$4.6 \\
After final DA cut & 112   &  106$\pm$2      & 17.0$\pm$0.3  & 47.5$\pm$1.1 & 39.2$\pm$0.7     &  0.5$\pm$0.05    &  1.5$\pm$1.0 \\
\hline
    \end{tabular}
    \caption[.]{
      \label{tab:vvqqsel}
             $\nu \bar{\nu} q \bar{q}$ channel: Data and simulation rates 
             after different steps of the analysis (see the description in 
             the text). The uncertainties result from the limited 
             simulation statistics. The CC 4-f background 
             refers to charged-current four-fermion processes such as 
             $W^+W^-$.}}    
  \end{center}
\end{table}

At both energies more events than predicted were observed before the final
selections. This resulted mainly from uncertainties in the description of 
the energy reconstruction of the $q\bar{q}(\gamma$) background. Unlike the 
other $ZZ$ decay channels studied, in which the full event energy is 
reconstructed most of the time, the selection of 
$ZZ \rightarrow \nu \bar{\nu} q \bar{q}$ events exploits the large missing 
energy characteristic of this channel, and is hence sensitive to the  
description of the low energy tail in the reconstruction of background 
processes such as $q\bar{q}(\gamma$). To study such effects and evaluate their
propagation to the final steps of the analysis, large statistics samples 
of $Z$ events, collected in the same conditions as the high 
energy data, were compared to the simulation 
to estimate corrections to the particle flow, in 
bins of momentum, polar angle and particle type. These consisted mainly 
of changes in multiplicities, to account for observed efficiency losses 
and possible duplication effects in the pattern recognition. The corrections, 
typically of one to a few percent in the barrel and endcaps respectively, 
were largest in the case of neutral particles and for high reconstructed 
momenta, and resulted in significantly improved agreement between data and 
simulation for energy flow observables such as the total charged and neutral 
energies, or the visible mass, including in the tails of these distributions. 
This procedure was then applied to all simulated high energy samples. 
The distribution of the discriminating variable obtained at 188.6~GeV 
after applying these corrections is shown in 
Figure~\ref{fig:vvqq_da_189_sys}b. 
The main effect was to increase the $q\bar{q}(\gamma$) background, 
by up to 15\%. 
The $ZZ$ cross-section fit was then repeated using this modified 
version of the 
simulation and differences were used as a conservative measure of the impact 
of uncertainties in the description of the low energy tail of the 
reconstruction. 
To reduce effects, regions where the $q\bar{q}(\gamma)$ background dominated 
were removed by a final   
cut on the discriminating variable at zero (indicated by an arrow in
Figure~\ref{fig:vvqq_da_189_sys}).
Above this cut, 22 events remained 
at 182.6 GeV and 112 events at 188.6 GeV. The systematic shift in the 
cross-section obtained amounted to 30\% relative to the expected value,
and was used to represent the systematic error at both energies. 

\section{Four-lepton final states\label{sub:LLLL}}
%
%
%

About 1\% of all $ZZ$ events lead to the $l^+l^-l^+l^-$ final state.
The event topology is clean and the only significant background comes 
from non-resonant $e^+e^-l^+l^-$ production. Events were selected if 
they contained between 4 and 8 charged particles, accompanied 
by at most 10 neutral particles, irrespective of particle 
identification.
In order to take into account final state radiation and bremsstrahlung 
effects for candidate electrons, the momenta of the charged particles 
were rescaled if the measured sum of energies of electromagnetic clusters 
in a narrow cone around the track direction was larger than the energy 
inferred from the track momentum measurement. The total invariant mass 
of the charged particles had to be greater than 50 GeV/c$^2$, and the 
minimum invariant mass after discarding any one of the charged particles 
larger than 20 GeV/c$^2$. All combinations of four charged particles with 
total charge zero were then examined, and a combination was selected if:

\begin{itemize}
\item
all four tracks had their impact parameters at the interaction point 
smaller than 3.0 and 0.5 cm, respectively in the projections containing 
the beam axis and perpendicular to it, and polar angles between 
$10^\circ$ and $170^\circ$,
\item
at least three of the four charged particles had momenta greater than 
5 GeV/c, and the least energetic particle a momentum greater than 2 GeV/c, 
\item
a system of two oppositely charged particles was found with both their
invariant and recoil masses within 10 GeV/c$^2$ of the $Z$ boson mass 
and having the same flavour (when both were identified as either 
electrons or muons), 
\item
the two particles complementary to this system were separated by at 
least $90^\circ$ from each other, 
\item
the invariant mass of all pairs of oppositely charged particles in the event 
exceeded 2~GeV/c$^2$.
\end{itemize}

At 182.6 GeV, no event was found, where predicted signal and background 
of respectively $0.098 \pm 0.005$ and $0.041 \pm 0.103$ events are expected.
At 188.6 GeV one event was observed, where predicted signal 
and background of respectively $0.683 \pm 0.022$ and $0.076 \pm 0.043$ events 
are expected.
The efficiencies of the selection were respectively $0.31 \pm 0.03$ and 
$0.41 \pm 0.02$ at each energy. 

To study potential systematic effects arising from the method based on mass 
windows used to disentangle the NC02 contribution from other non-resonant 
four fermion processes, the results were checked with a different signal 
definition, consisting of generating two $l^+l^-l^+l^-$ samples, the first 
using NC02 graphs only and the second using all other tree-level graphs, 
with NC02 excluded. The first sample was used to define the signal and the
second was used to estimate the background from other four-fermion processes.
Results obtained in this way were fully consistent with those obtained using 
the mass window method.

\section{Two isolated leptons with missing energy}
%
%
%

Events from $ZZ \to \nu \bar{\nu} \mu^+ \mu^- , \nu \bar{\nu} e^+ e^- $ 
processes are characterized by two relatively acollinear charged leptons 
of the same flavour, with both invariant and recoil masses close to that
of the $Z$, and by large missing energy. Although it has different production
kinematics, the $WW$ process also contributes to these final states with a 
large cross-section. A significant fraction of the corresponding events have 
exactly the same features and constitute a dominant background.

To ensure good reconstruction, tracks were required to have impact parameters 
at the interaction point smaller than 3.0 and 0.5 cm, respectively in the 
projections containing the beam axis and perpendicular to it, and polar angles 
between $20^\circ$ and $160^\circ$. As for the $l^+l^-l^+l^-$ channel, final 
state radiation and bremsstrahlung effects for candidate electrons were taken 
into account by rescaling the momenta of the charged particles if the measured 
sum of energies of electromagnetic clusters in a narrow cone around the track 
direction was larger than the energy inferred from the track momentum measurement.
Events with two particles identified as $e^+e^-$ or $\mu^+\mu^-$ were selected if 
their total energy was less than 60\% of that of the centre-of-mass, if the angle 
between them was in the range {$140^\circ$-$170^\circ$}, if the polar angle of 
the missing momentum vector was between $25^\circ$ and $155^\circ$, and if the 
reconstructed invariant masses satisfied:

min\{$|~M_{Z} - m(l^+l^-)~|~$,$~|~M_{Z} - m_{recoil}(l^+l^-)~|$\} 
      $<$ 4 GeV/c$^2$, and

max\{$|~M_{Z} - m(l^+l^-)~|~$,$~|~M_{Z} - m_{recoil}(l^+l^-)~|$\} 
      $<$ 8 GeV/c$^2$,

\noindent 
where $m_{recoil}(l^+l^-)$ is the invariant mass recoiling 
against the $l^+l^-$ pair.

At 182.6 GeV, no event was found, with predicted signal and background 
of $0.109 \pm 0.007$ and $0.348 \pm 0.155$ events respectively.
At 188.6 GeV, two events were observed, with predicted signal and background 
of $0.839 \pm 0.028$ and $1.025 \pm 0.263$ events respectively.
The efficiencies of the selection were $0.26 \pm 0.02$ and 
$0.30 \pm 0.02$ at each energy respectively.

The main systematic effects in this channel were from uncertainties in the 
lepton identification efficiencies, both of which are at the few per cent 
level, and to a lesser extent from uncertainties in the $WW$ cross-section 
used. In view of the limited statistics in this channel, these effects were 
neglected.


%
%
%
\section{Combined NC02 cross-section}
%
%
%

The cross-sections in the $q \bar{q} q \bar{q}$ and $\nu \bar{\nu} q \bar{q}$ 
channels were determined individually by means of binned maximum likelihood 
fits to the distributions of the combined variable defined in each of the 
corresponding analyses. To derive a combined value for the NC02 cross-section 
from all the 
six channels analysed, a global likelihood was constructed by combining 
the likelihoods from the fits performed in these two channels with the Poisson 
probabilities for observing the number of events seen in the four other 
channels ($\mu^+ \mu^- q \bar{q}$, $e^+e^- q \bar{q}$, $l^+l^-l^+l^-$, and 
$\nu \bar{\nu} l^+l^-$), given the predicted numbers. This global likelihood 
was 
then maximised with respect to variations of the value of the NC02 
cross-section, 
assuming branching ratios of the $Z$ fixed to those expected in the 
Standard Model, 
and taking into account the scaling factors listed in 
table~\ref{table:scalingfac}. 

The impact of the systematic uncertainties affecting the different 
final states was studied by repeating the fits with modified assumptions 
on backgrounds and efficiencies (see the corresponding sections). 
Statistical errors exceeded the estimated systematic ones in each 
of the channels taken separately. To estimate the total systematic 
error on the combined result, the different uncertainties were assumed, 
conservatively, to be correlated, and their effects were added. This 
resulted in a total error of $\pm$11\% relative to the expected value, 
dominated by the uncertainty in the $\nu \bar{\nu} q \bar{q}$ channel.

The values for the NC02 cross-sections obtained were:

\begin{tabbing}
~~~~~~~~~~\=$ \sigma_{\tiny NC02}$~(182.6 GeV) 
          \== 0.38 $ \pm $ $0.18 ~(stat)$ $ \pm $ $0.04 ~(syst)$ pb, \\
~~\\
          \>$ \sigma_{\tiny NC02}$~(188.6   GeV) 
          \>= 0.60 $ \pm $ $0.13 ~(stat) \pm 0.07~(syst)$ pb.
\end{tabbing}

%
%
%
%
%
%



\begin{figure}[tbh]
  \begin{center}
    \mbox{\epsfxsize=17.0cm\epsfysize=17.0cm\epsffile{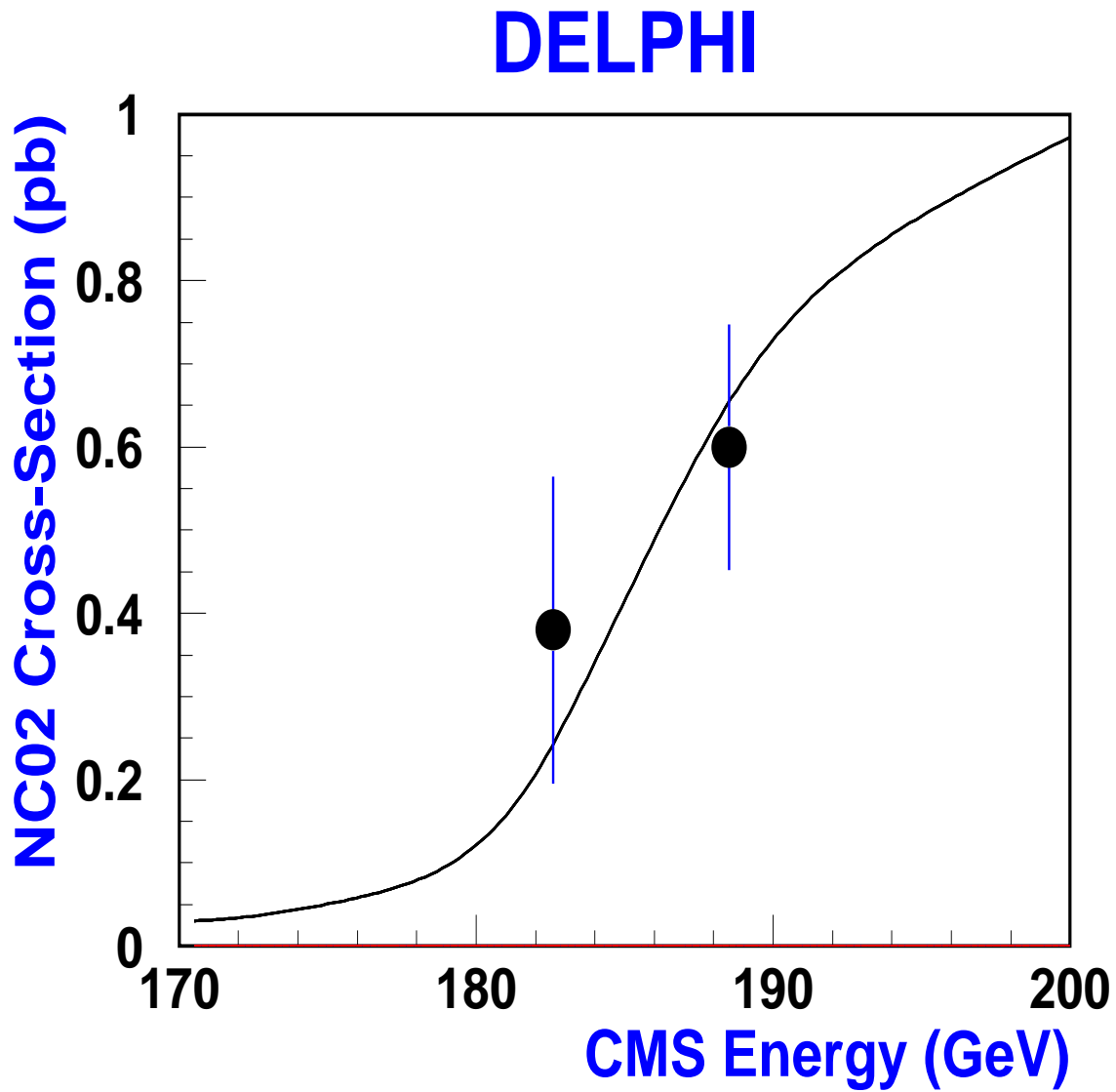}}    
  \caption{ Combined NC02 cross-sections measured from data collected in
            1997 and 1998. The solid curve was computed
            using the {\tt YFSZZ} Monte-Carlo ~\cite{lancon}.
            Values obtained with the {\tt EXCALIBUR} Monte-Carlo were 
            within $\pm 1$ \% of those obtained with {\tt YFSZZ}.
           \label{fig:zzcro189}}
  \end{center}
\end{figure}

These values are consistent with the Standard Model expectations of 
0.25 pb and 0.65 pb at each energy, respectively, predicted by  
{\tt EXCALIBUR}. The agreement is illustrated in Figure~\ref{fig:zzcro189}.
Measurements of on-shell $ZZ$ production by the three other LEP 
collaborations can be found in ~\cite{l3zz,alephzz,opalzz}.


\subsection*{Acknowledgements}
\vskip 3 mm
 We are greatly indebted to our technical 
collaborators, to the members of the CERN-SL Division for the excellent 
performance of the LEP collider, and to the funding agencies for their
support in building and operating the DELPHI detector.\\
We acknowledge in particular the support of \\
Austrian Federal Ministry of Science and Traffics, GZ 616.364/2-III/2a/98, \\
FNRS--FWO, Flanders Institute to encourage scientific and technological 
research in the industry (IWT), Belgium,  \\
FINEP, CNPq, CAPES, FUJB and FAPERJ, Brazil, \\
Czech Ministry of Industry and Trade, GA CR 202/96/0450 and GA AVCR A1010521,\\
Danish Natural Research Council, \\
Commission of the European Communities (DG XII), \\
Direction des Sciences de la Mati$\grave{\mbox{\rm e}}$re, CEA, France, \\
Bundesministerium f$\ddot{\mbox{\rm u}}$r Bildung, Wissenschaft, Forschung 
und Technologie, Germany,\\
General Secretariat for Research and Technology, Greece, \\
National Science Foundation (NWO) and Foundation for Research on Matter (FOM),
The Netherlands, \\
Norwegian Research Council,  \\
State Committee for Scientific Research, Poland, 2P03B06015, 2P03B11116 and
SPUB/P03/DZ3/99, \\
JNICT--Junta Nacional de Investiga\c{c}\~{a}o Cient\'{\i}fica 
e Tecnol$\acute{\mbox{\rm o}}$gica, Portugal, \\
Vedecka grantova agentura MS SR, Slovakia, Nr. 95/5195/134, \\
Ministry of Science and Technology of the Republic of Slovenia, \\
CICYT, Spain, AEN96--1661 and AEN96-1681,  \\
The Swedish Natural Science Research Council,      \\
Particle Physics and Astronomy Research Council, UK, \\
Department of Energy, USA, DE--FG02--94ER40817, \\

\end{document}